\documentclass[12pt]{iopart}

\usepackage{iopams}  
 \usepackage{graphics}
 \usepackage{graphicx}

\begin{document}

\title[Secondary Magnetic Reconnection Driven by Kink Instability]{Signatures of Secondary Collisionless Magnetic Reconnection Driven by Kink Instability of a Flux Rope}

\author{S. Markidis$^1$, G. Lapenta$^2$, G.L. Delzanno$^3$, P. Henri$^4$ , M.V. Goldman$^5$, D.L. Newman$^5$, T. Intrator$^3$,  E. Laure$^1$}

\address{$^1$ High Performance Computing and Visualization (HPCViz) Department, 
KTH Royal Institute of Technology,
Stockholm, Sweden \\
$^2$ Center for mathematical Plasma Astrophysics (CmPA), KULeuven, Leuven, Belgium \\
$^3$ Los Alamos National Laboratory, Los Alamos, NM USA\\
$^4$ LPC2E, CNRS, Orl\'eans, France\\
$^5$ Department of Physics and CIPS, University of Colorado, Boulder, CO USA\\
}
\ead{markidis@kth.se}
\begin{abstract}
The kinetic features of secondary magnetic reconnection in a single flux rope undergoing internal kink instability are studied by means of three-dimensional Particle-in-Cell simulations. Several signatures of secondary magnetic reconnection are identified in the plane perpendicular to the flux rope: a quadrupolar electron and ion density structure and a bipolar Hall magnetic field develop in proximity of the reconnection region. The most intense electric fields form perpendicularly to the local magnetic field, and a reconnection electric field is identified in the plane perpendicular to the flux rope. An electron current develops along the reconnection line in the opposite direction of the electron current supporting the flux rope magnetic field structure. Along the reconnection line, several bipolar structures of the electric field parallel to the magnetic field occur making the magnetic reconnection region turbulent. The reported signatures of secondary magnetic reconnection can help to localize magnetic reconnection events in space, astrophysical and fusion plasmas.

\end{abstract}

\maketitle

\section{Introduction}
Flux ropes are bundles of magnetic field lines wrapped around an axis. They are ubiquitous structures in space, astrophysical and fusion device plasmas. The presence of flux ropes has been observed in planetary magnetotail \cite{Eastwood:2005}, dayside magnetopause \cite{Paschmann:1979}, and in solar corona \cite{Gosling:1990}. Astrophysical jets can be treated as flux ropes \cite{lovelace1976dynamo,li2006modeling,lapenta2006kink}. Flux ropes can also reproduce to a first approximation the magnetic field geometry of fusion devices such as Z-pinch or Tokamak fusion reactors \cite{freidberg:2007}. The kink instability is one of the most studied instabilities of flux ropes \cite{ara1978magnetic}. This instability is particularly important in fusion devices, where it can disrupt the flux rope limiting the confinement of plasma \cite{von1974studies}. Because of this, many analytical and numerical studies have been devoted to the understanding of its dynamics. The kink instability has been extensively studied using MHD modeling. It has been examined with Particle-in-Cell methods \cite{MarkidisEPS,Restante:2013,mishchenko:2012} only recently to enlighten possible kinetic effects during the evolution of the kink instability.

This paper presents a simulation study of secondary magnetic reconnection triggered by the internal kink instability of a single flux rope in a periodic domain along the axis direction. Secondary magnetic reconnection occurs when magnetic reconnection is driven by a primary instability that forces the magnetic field lines from different disconnected magnetic regions to be in contact and reconnect. Therefore, secondary magnetic reconnection develops over a dynamic time scale imposed by the primary instability that acts as a driver. For instance, secondary magnetic reconnection might be triggered by Kelvin-Helmholtz (KH) instability \cite{HenriNEW}. The vortices created by the KH instability force the magnetic field lines from different magnetic domains to reconnect. An additional example is the magnetic reconnection driven by interchange instability affecting the reconnection fronts in magnetotail magnetic reconnection \cite{Vapirev:2013, LapentaNEW2}. Secondary magnetic reconnection has an important role in the overall dynamics of the primary instability: it converts magnetic field energy to kinetic energy of the plasma allowing for a local reconfiguration of the magnetic field topology.

Several signatures of magnetic reconnection have been identified in simplified configurations using computer simulations. The most famous example of such configurations is the Harris current-sheet model \cite{harris1962plasma}. The vast majority of simulations starts from the Harris equilibrium. In this configuration, the most famous signature of magnetic reconnection is the quadrupolar structure of the Hall magnetic field developing on the reconnection plane \cite{sonnerup1981evidence,Rogers:2003}. The quadrupolar structure of the Hall magnetic field originates because of the decoupling of electron and ion dynamics in a small region in proximity of the reconnection line. Additional signatures of magnetic reconnection are the presence of regions of space with depleted density (density cavities)  \cite{markidis2012three}, of bipolar electric field structures and electron holes in the phase space \cite{LapentaGRL:2011,Lapenta:2010}, of whistler and kinetic Alfv\'en waves \cite{GoldmanNEW, LapentaNEW}. In this paper, we focus on signatures of magnetic reconnection in a more complicated initial configuration, such as the single flux rope set-up. In this configuration, we found signatures that are similar to the ones of magnetic reconnection in the Harris sheet \cite{Rogers:2003} and some new features characteristic of magnetic reconnection during the kink instability.

The main new contribution of this paper is to identify the features of magnetic reconnection driven by internal kink instability. These features can be used by scientists to determine reconnection sites developing during the kink instability of a single flux rope and to increase the understanding of the complex phenomena involved in the kink instability of a flux rope.

The paper is organized as follows. The simulation model, the initial configuration of the flux rope and the parameters in use are presented in Section 2. The main features and the signatures of secondary magnetic reconnection are presented in Section 3. Section 4 discusses the results, summarizes the results and finally outlines possible future extension of this work.

\section{Simulation Model}
We present three-dimensional Particle-in-Cell (PIC) simulations of a flux rope instability. By using a PIC model, we describe correctly the kinetic behavior of the flux rope plasma. The internal kink instability develops over time scale of hundred of ion gyro-periods. These time scales are much larger than the typical time scales that can be covered by standard PIC methods. In order to cover such a large period of time, we use an implicit PIC approach \cite{Lapenta:2006} to avoid numerical instability when using large simulation time steps.

The single flux rope is modeled with a simple screw-pinch configuration \cite{freidberg:2007} as in previous works \cite{lapenta2006kink,Restante:2013}. In this configuration, the initial magnetic field in cylindrical coordinates is:
 \begin{equation}
 \label{InitB}
\left\{
\begin{array}{l}
B_{\theta} = \frac{B_0 r}{r^2 + w^2} \\
B_{z} = B_0 \\
B_{r} = 0
\end{array}
\right.
\end{equation}
The magnetic field lines are helical with a pitch determined by the amplitude $B_0 = 0.1  \ m_i c \omega_{pi} / e$, where $m_i$ is the ion mass, $c$ is the speed of light in the vacuum, $\omega_{pi}$ is the ion plasma frequency and $c$ the elementary charge. The parameter $w$ is chosen to be $1.0 \  d_i$ ($d_i= c/\omega_{pi}$ is the ion skin depth). The density $n_0 = 1$ is initially uniform. The initial current is computed numerically by solving the Ampere's law with the given magnetic screw-pinch configuration. In our set-up, only electrons carry out the initial current. The electron drift velocity of the computational particles is calculated from the electron current as $\mathbf{V_{e0}} = \mathbf{J_{e0}}/(e n_0)$.  Ions do not have an initial drift velocity. The pressure is calculated by imposing a force balance in the domain. The thermal velocity of the computational particles is computed from the local value of the pressure. In this configuration, the plasma beta, the ratio of plasma pressure to the magnetic pressure, is $\beta \approx 0.2$. The electron beta is $\beta_e \approx 0.02$.

The simulation box is $L_x \times L_y \times L_z= 4 \pi  \ d_i\times 4 \pi \  d_i \times 8 \pi \  d_i$ long and discretized in a Cartesian domain with $n_x \times n_y \times n_z= 128 \times 128  \times 256$ points. An artificial ion to electron mass ratio equal to 25 is chosen to reduce the simulation execution time. The grid spacing results $\Delta x = \Delta y = \Delta z = 0.098 \ d_i = 0.78 \ d_e$, where $d_e$ is the electron skin depth. Because the electron skin depth is not fully resolved by the grid spacing, the proposed simulation does not model the dissipative kink mode (the collisionless analogous of the resistive kink mode) but the kink mode under study is an ion-kinetic modification of the ideal kink mode \cite{pegoraro1989internal}. Perfect conductor boundary conditions for the electromagnetic field and reflecting for particles are used in the $x$ and $y$ directions. The system is periodic in the $z$ direction, and therefore it can be considered infinite in that direction. The simulation time step is $\Delta t = 0.3 \   \omega_{pi} ^{-1}$. In this configuration, $\Delta t =   0.3 \ \omega_{pi} ^{-1} = 2.4 \ \omega_{p} ^{-1}$, where $\omega_{p}$ is the plasma frequency. The time step in use is larger than the maximum time step ($2 \ \omega_{p} ^{-1}$) allowed in standard explicit PIC methods. A total of 226 million particles are used. Simulations are carried out with the parallel fully kinetic and fully electromagnetic PIC code iPIC3D  \cite{Markidis:2010} on 512 cores of the Lindgren supercomputer at KTH Royal Institute of Technology. In our simulations, we use the core magnetic field $B_0$ to define the ion cyclo-frequency $\Omega_{ci} =  B_0 e/m_e$, and Alfv\'en velocity $V_A = B_0/ \sqrt{4 \pi e n_0}$. In this configuration, $c/V_A = \omega_{pi}/ \Omega_{ci} $ is 10. The ion and electron gyro-radii are respectively $0.4 \ d_i$ and  $0.03 \ d_i$.	 In this set-up, the time step and grid the spacing resolve all the time and space scales relevant to the study of magnetic reconnection. A relatively high number of computational particles (if compared with standard PIC simulations) has been chosen to decrease the numerical noise arising from the particle description of plasma. It is therefore reasonable to expect that severe numerical artifacts do not affect the simulation of magnetic reconnection.

This initial configuration is unstable against the kink instability. In fact, the safety factor $q = 2 \pi r B_z / B_\theta L_z $ is smaller than one in a region of space in the computational domain, and therefore the system results unstable against kink instability \cite{kadomtsev1958book}. The surface $q = 1$ is called "resonant surface" and magnetic reconnection will develop along the resonant surface when the kink instability reaches the non linear stage \cite{kadomtsev1958book}. The resonant surface $r_s$ is located at $r \approx 1.7 \  d_i$ in our simulation set-up.

\section{Results}
The evolution of the kink instability and of the secondary magnetic reconnection in a three-dimensional PIC simulation is presented. Figure 1 shows an isocontour of the axial component of the electron current at four successive times $t=0, 60 \ \Omega_{ci}^{-1}, 75 \ \Omega_{ci}^{-1}, 90 \ \Omega_{ci}^{-1}$. The electron current is represented as a vertical tube in the plot of axial component ($z$ direction) of the electron current $J_{ez}$. The electron current tube starts bending approximately at time $t= 75 \ \Omega_{ci}^{-1}$, and the kink instability eventually fully develops at time $t= 90 \ \Omega_{ci}^{-1}$.

\begin{figure}[ht]
\includegraphics[width=0.95 \columnwidth]{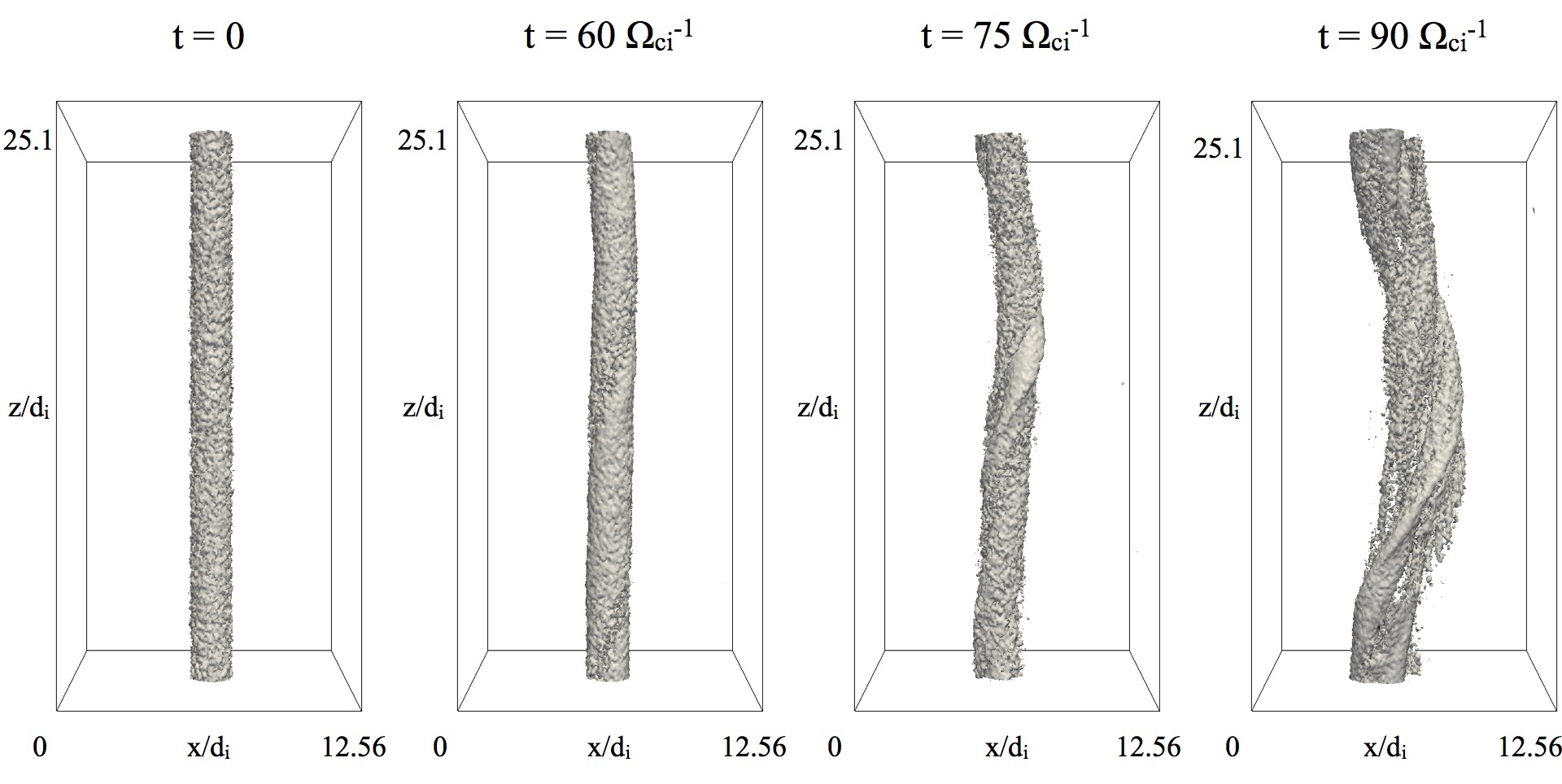}
\caption{Isocontour of the axial component of the electron current $J_{ez} =   \ 0.005 e n_0 c $ at times $t=0, 60 \ \Omega_{ci}^{-1}, 75 \ \Omega_{ci}^{-1}, 90 \ \Omega_{ci}^{-1}$. A kink of the electron channel develops approximately after $t=0 75 \ \Omega_{ci}^{-1}$.}
\label{Jz}
\end{figure}

Figure 2 shows the magnetic field lines as grey tubes at times $t=0, 84 \ \Omega_{ci}^{-1}, 105 \ \Omega_{ci}^{-1}, 120 \ \Omega_{ci}^{-1}$. The magnetic field lines progressively untwist as effect of magnetic reconnection \cite{kadomtsev1958book}.
\begin{figure}[ht]
\includegraphics[width=0.95 \columnwidth]{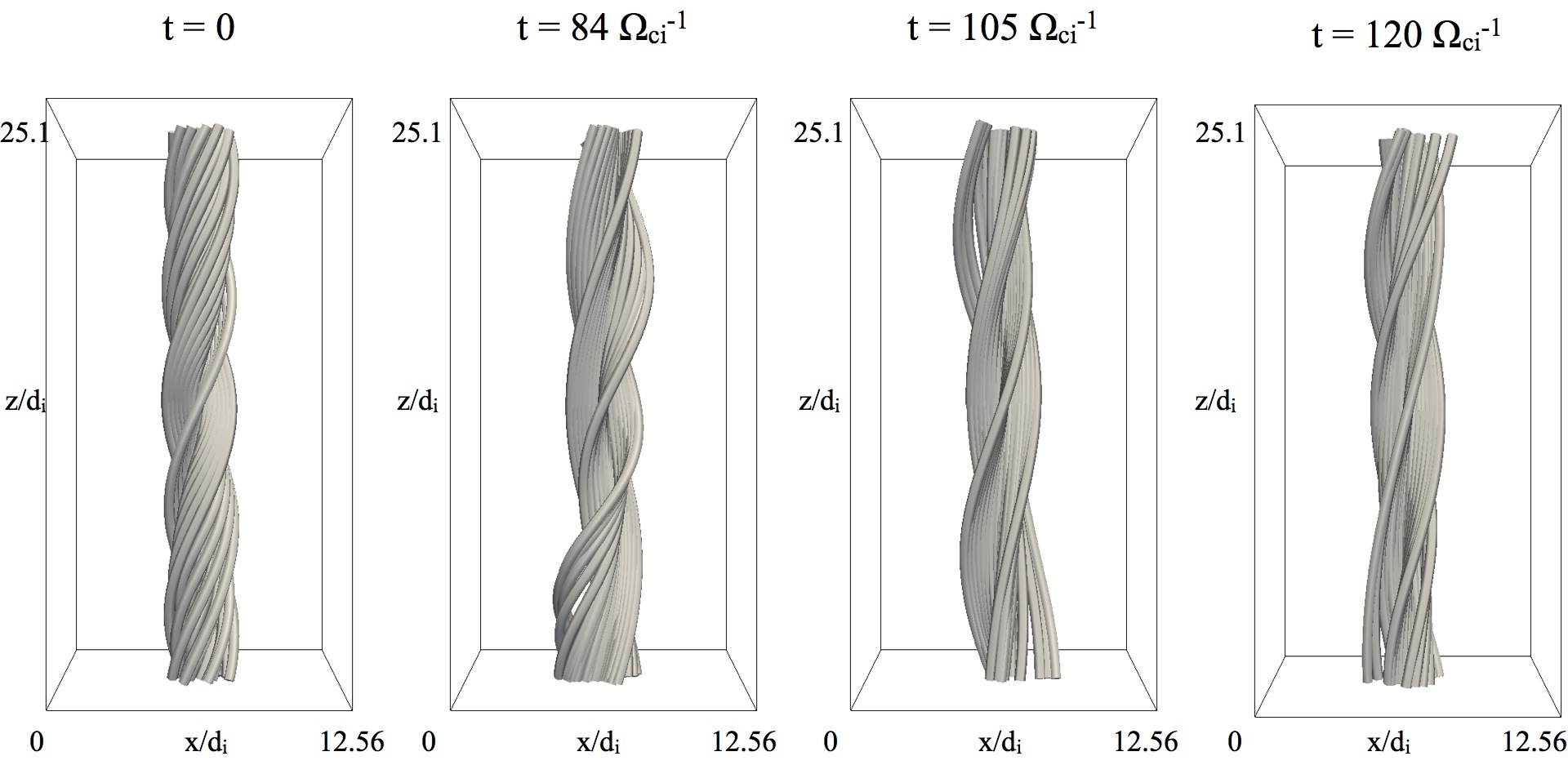}
\caption{Magnetic field lines at times $t=0, 84 \ \Omega_{ci}^{-1}, 105 \ \Omega_{ci}^{-1}, 120 \ \Omega_{ci}^{-1}$. The magnetic field lines untwist as macroscopic effect of magnetic reconnection.}
\label{Blines}
\end{figure}

We focus on the secondary magnetic reconnection signatures in the plane $(x,y)$ perpendicular to the axis of the flux rope. A convenient quantity to identify magnetic reconnection is the auxiliary magnetic field $\mathbf{B}^*$, defined as $\mathbf{B}^* = \mathbf{B}_\theta - (2r/L_x)  B_z$. Figure 3 shows $\mathbf{B}^*$ at $t= 0, 81, 87 \ \Omega_{ci}^{-1}$ in the plane $z = L_z/2, r < 3.5 \ d_i$. Initially,  the quiver plot shows two disconnected $\mathbf{B}^*$ regions. In these two regions, $\mathbf{B}^*$ points at opposite directions. $\mathbf{B}^*$ vanishes at $r=r_s$ represented as red dashed line in Figure 3. As the simulation progresses, the internal $\mathbf{B}^*$ region moves outward and crosses the resonant surface as effect of the kink instability of the flux rope. This is clear by inspecting $\mathbf{B}^*$ configuration at time $t=81 \ \Omega_{ci}^{-1}$ in Figure 3. As the internal $\mathbf{B}^*$ region crosses the resonant surface, magnetic reconnection occurs, forming two new $\mathbf{B}^*$ regions. We have found that $\mathbf{B}^*$ is almost independent on $z$. For this reason, we focus on studying secondary magnetic reconnection in the plane $z = L_z/2$. Almost identical features can be found on different $z$ planes.
\begin{figure}[ht]
\includegraphics[width=0.95 \columnwidth]{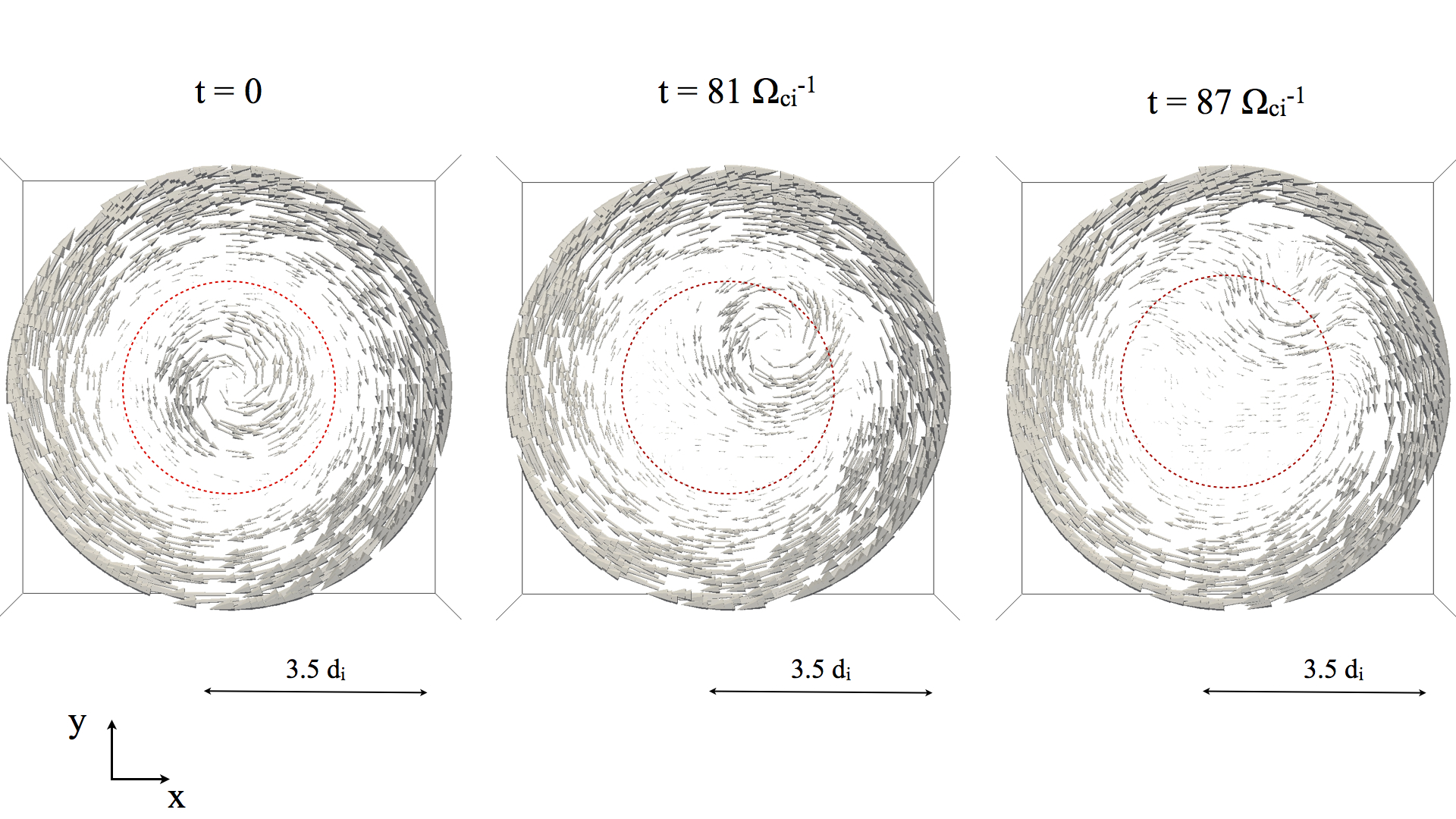}
\caption{Quiver plot of $\mathbf{B}^*$ at time $t=0$, $t=81 \ \Omega_{ci}^{-1}$  and $t=87 \ \Omega_{ci}^{-1}$ in the plane $z = L_z/2, r < 3.5 \ d_i$.  Initially, the quiver plot shows two disconnected $\mathbf{B}^*$ regions ($\mathbf{B}^*$ pointing at different directions). At time $t=81 \ \Omega_{ci}^{-1}$, the $\mathbf{B}^*$ internal region moves outward as effect of the kink instability. At time $t=87 \ \Omega_{ci}^{-1}$, the two initially disconnected $\mathbf{B}^*$ regions reconnect. The resonant surface is shown with dashed red line.}
\label{Bstar}
\end{figure}

The plasma density is initially uniform. However, density inhomogeneities develop under the effect of secondary magnetic reconnection. Figure 4 shows the contour plot of the electron (panel a) and ion (panel b) densities in the plane $z = L_z/2, r < 3.5 \ d_i$ at time $t = 84 \  \Omega_{ci}^{-1}$. The densities are normalized to the initial density $n_0$. The formation of density cavities is a characteristic of magnetic reconnection. They are localized along the separatrices and they are crossed by intense electron jets. It is found that the regions with enhanced and depleted densities form a quadrupolar structure, as outlined with dashed lines in Figure 4 panel b). The maximum and minimum densities are 50 \% peak initial density. Density cavities are approximately $4 \ d_{i}$ long and  $1 \ d_{i}$ thick. Magnetic reconnection occurs in a region of space with density $n \approx n_0/2$. Previous two-dimensional simulations \cite{markidis2011kinetic} show that grid resolution in the proposed simulation allows to capture all the relevant dynamics for studying magnetic reconnection.

\begin{figure}[ht]
\includegraphics[width=1 \columnwidth]{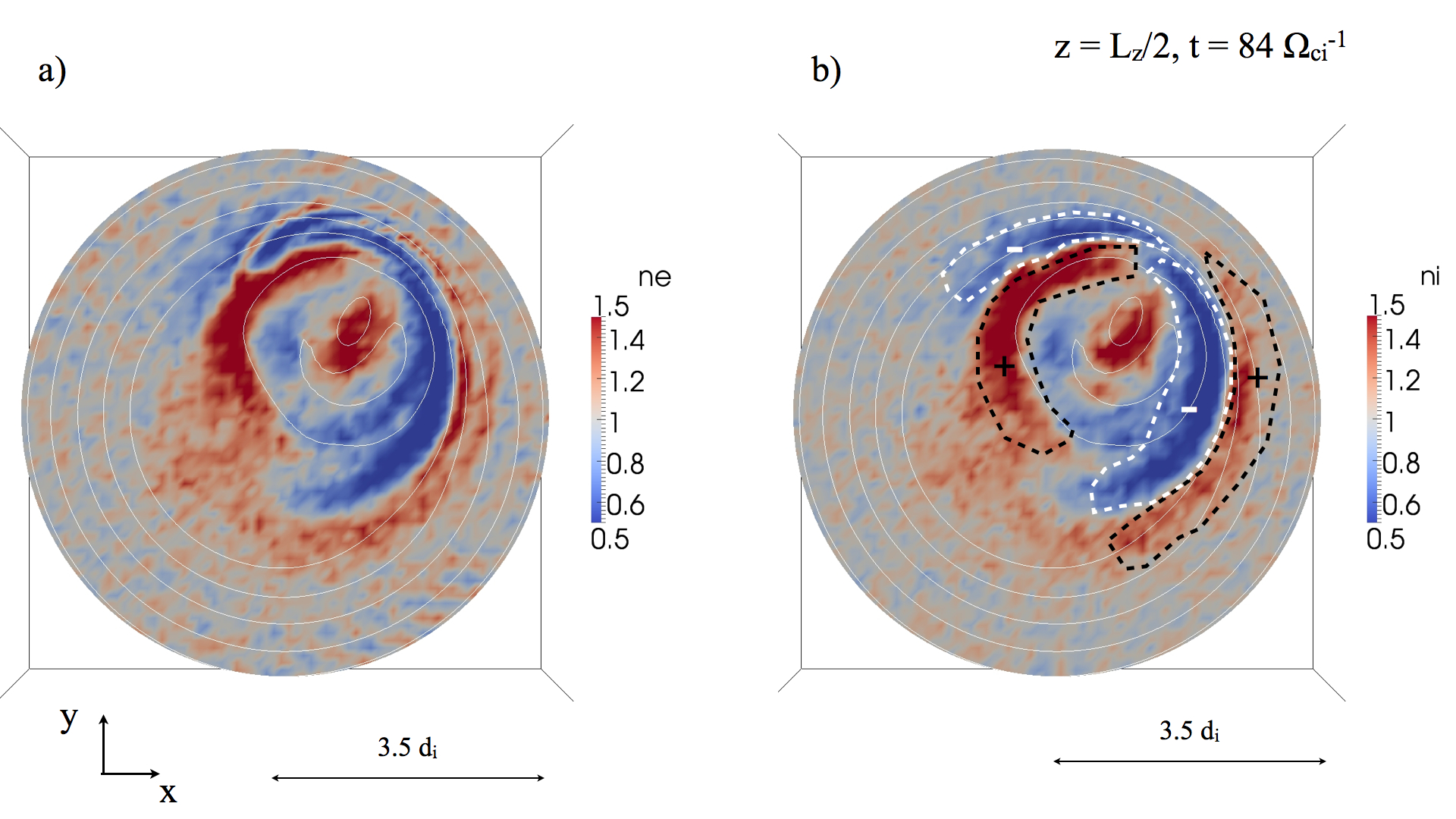}
\caption{Isocontour plots of the electron (panel a) and ion (panel b) densities in the plane $z = L_z/2, r < 3.5 \ d_i$ at time $t = 84 \ \Omega_{ci}^{-1}$. The densities are normalized to $n_0$. The densities plot shows region of enhanced and depleted densities (organized as quadrupolar structure) in proximity of the reconnection region. The dashed lines shows the quadrupolar structure in panel b). The white line represent $\mathbf{B}^*$ contour-lines.}
\label{neni}
\end{figure}

Initially in the simulation, the current is primarily along the flux rope axis ($z$ direction) and carried out by electrons while ions have only an initial thermal velocity and no drift velocity. Panel a) of Figure 5 shows a contour plot of the $z$ component of the electron bulk velocity $\mathbf{V}_e = \mathbf{J}_e/\rho_e$, where velocities are normalized to the Alfv\'en velocity $V_A$. At the center of the contour plot, the initial electron current is directed in negative $z$ direction and it can be still seen in blue color. In the proximity of the reconnection site, a thin layer of electrons flowing in the opposite direction of the initial electron current can be observed in the red color. This current is localized in the region characterized by a relatively low electron density (Figure 4).  The electron peak bulk velocity is approximately $1.5 \ V_A$ at center of the flux rope, while an opposite flow at $- 1.5 \ V_A$ develops in the low density region of the flux rope. Panel b) of Figure 5 shows a contour plot of the $z$ component of the electron bulk velocity $\mathbf{V}_i = \mathbf{J}_i/\rho_i$. This shows a quadrupolar structure, similar to high and low density structures in Figure 4. The ion peak bulk velocity is approximately $0.2 \ V_A$, approximately eight times smaller than the electron peak bulk velocity.
\begin{figure}[ht]
\includegraphics[width=1 \columnwidth]{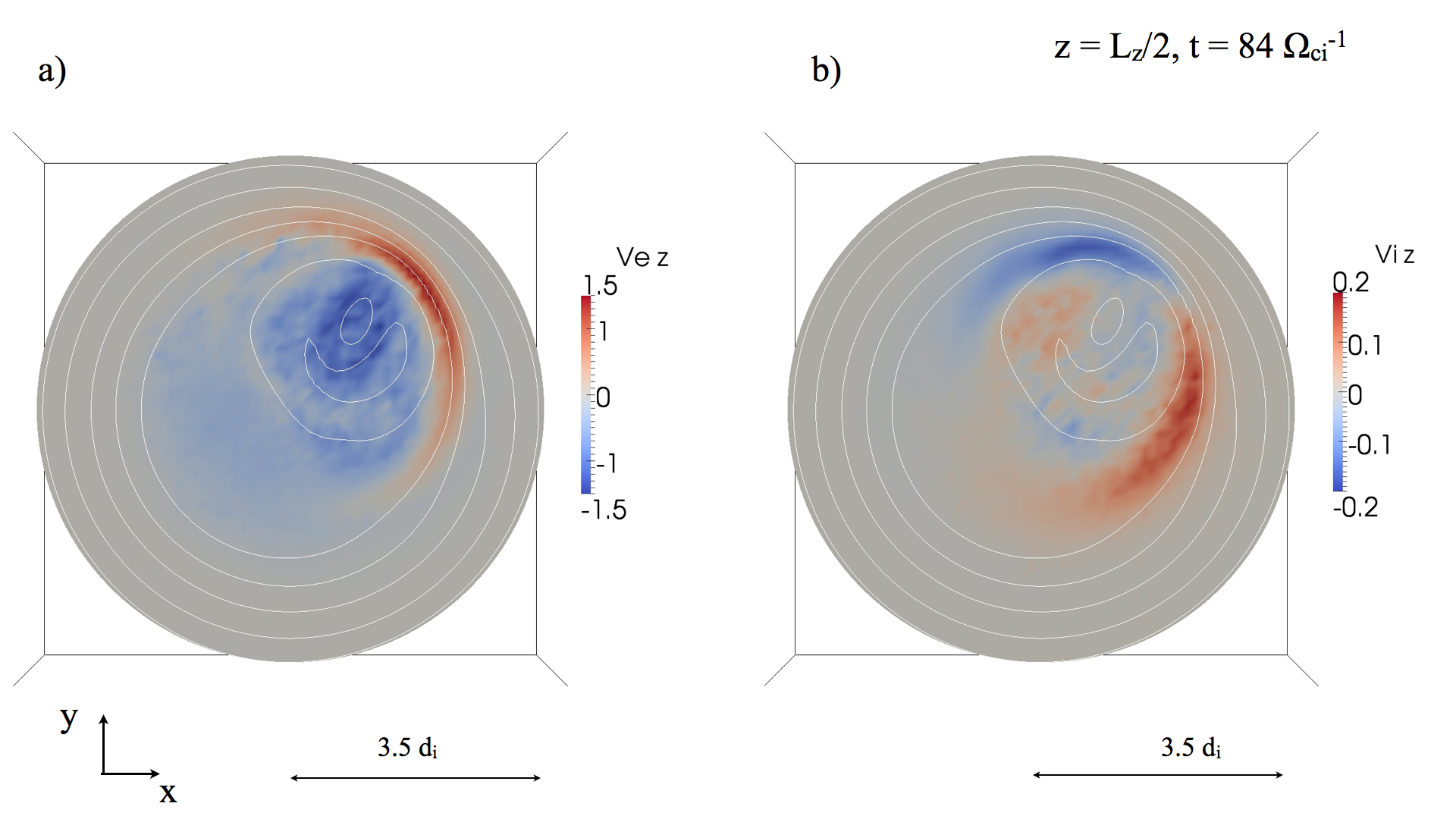}
\caption{Isocontour plot of the $z$ component of the electron and ion bulk velocities $V_{ez}, V_{iz}$ in the plane $z = L_z/2, r < 3.5 \ d_i$ in panels a) and b). Velocities are normalized to Alfv\'en velocity $V_A$. The white lines represent $\mathbf{B}^*$ contour-lines.}
\label{Velocities}
\end{figure}

The panel a) of Figure 6 shows an isocontour plot of the electric field intensity with a superimposed quiver in the plane $z = L_z/2, r < 3.5 \ d_i$ at time $t = 84 \ \Omega_{ci}^{-1}$. The electric field is normalized to $B_0 V_A / c$.  The electric field has the strongest components in the plane $(x,y)$ and is localized in proximity of the density cavity regions. A weaker component of the electric field in the $z$ direction is observed also. A contour plot of the $z$ component of the electric field in the plane $z = L_z/2, r < 3.5 \ d_i$ at time $t = 84 \ \Omega_{ci}^{-1}$ is shown in panel b) of Figure 6. Its peak value is  $0.1 \  B_0 V_A / c$.

\begin{figure}[ht]
\includegraphics[width=1 \columnwidth]{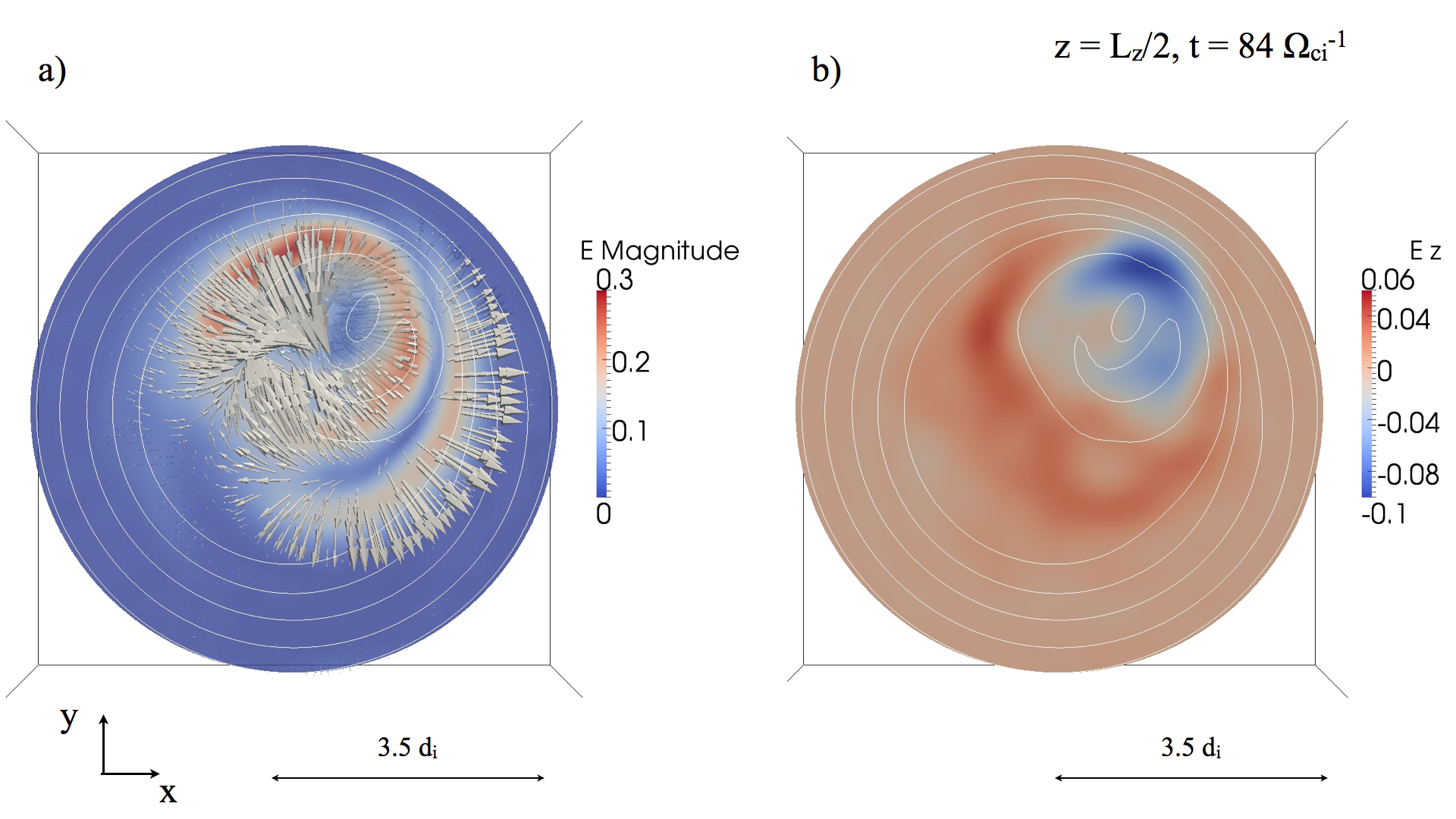}
\caption{Isocontour plot of the electric field intensity with a superimposed $\mathbf{E}$ quiver plot in the plane $z = L_z/2, r < 3.5 \ d_i$  at time $t = 84 \ \Omega_{ci}^{-1}$ is shown in panel a). Isocontour plot of the $z$ component of the electric field is shown in panel b). The electric field is normalized to $B_0 V_A / c$. The white lines represent $\mathbf{B}^*$ contour-lines.}
\label{Efield}
\end{figure}

Figure 7 presents an isocontour plot of the magnetic field intensity and a quiver plot of magnetic field at time $t = 84 \ \Omega_{ci}^{-1}$ in panel a). The magnetic field values are normalized to $B_0$. The magnetic field is dominated by the core magnetic field in the $z$ direction and its value is close to $B_0$ in the region $r < 3.5 \ d_i$. Panel b) shows the $z$ component of the magnetic field. A bipolar structure of the $B_z$ component has been found in proximity of the reconnection region. The peak value of the $B_z - B_0$ is approximately $0.2 \  B_0$.

\begin{figure}[ht]
\includegraphics[width=1 \columnwidth]{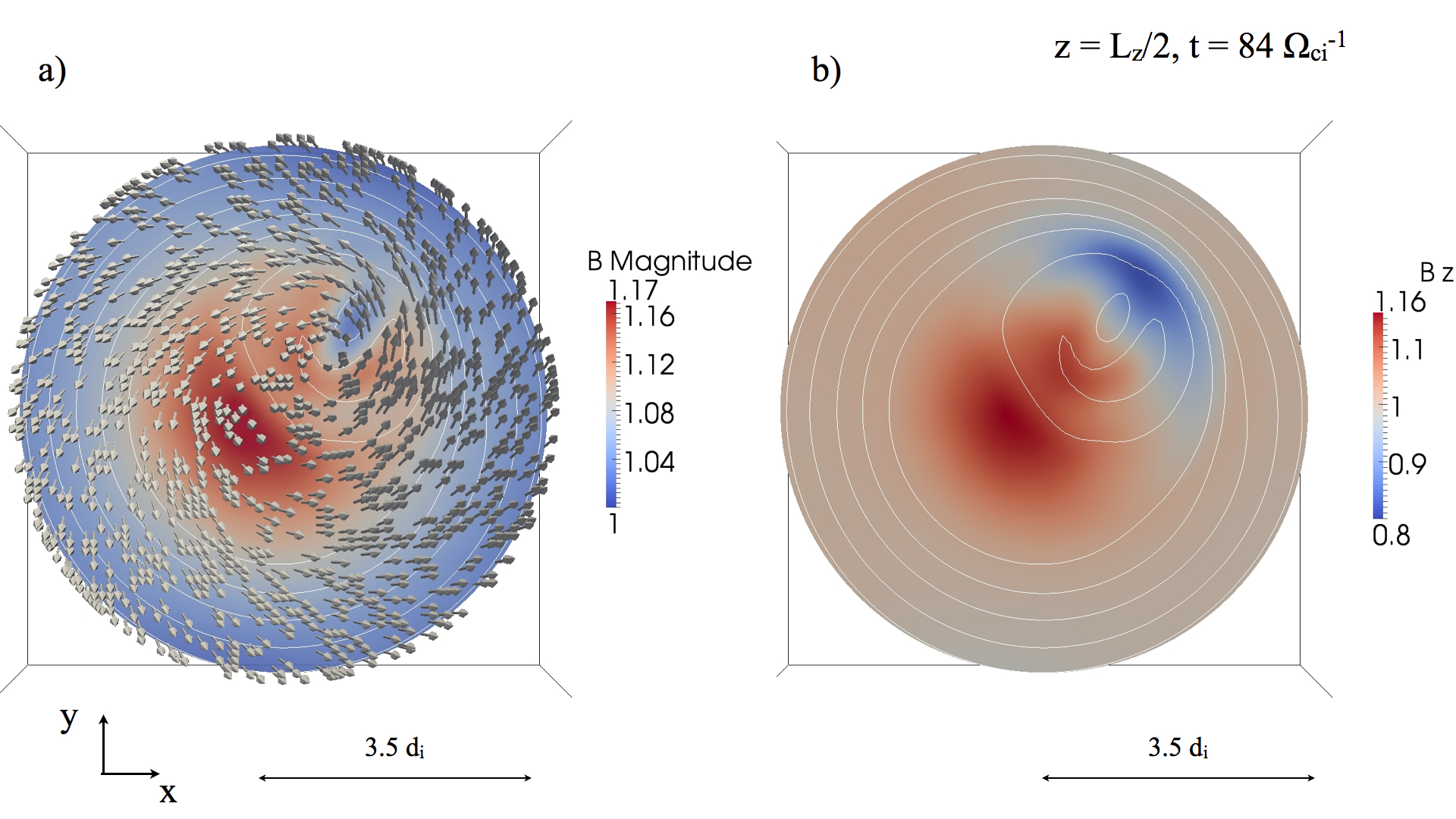}
\caption{Isocontour plot of the magnetic field intensity with a superimposed quiver plot in the plane $z = L_z/2, r < 3.5 \ d_i$  at time $t = 84 \ \Omega_{ci}^{-1}$ is shown in panel a). Isocontour plot of the $z$ component of the magnetic field is shown in panel b). The magnetic field values are normalized to core magnetic field $B_0$. The white lines represent $\mathbf{B}^*$ contour-lines.}
\label{Jz}
\end{figure}

As seen in Figure 5, two electron currents develop in opposite directions during secondary magnetic reconnection. Figure 8 shows a three dimensional contour plot for axial component of the electron current $J_{ez} = 0.005 \ e n_0 c$ (red color) and $J_{ez} = - 0.005 \ e n_0 c$. Initially the electron current component in the $z$ direction is only positive, while later on an electron current develops in the negative $z$ direction. At time $t = 90 \ \Omega_{ci}^{-1}$, the electron current is perturbed. This might suggest that an electron-scale streaming instability could occur at the interface of the two opposite currents.

\begin{figure}[ht]
\includegraphics[width=1 \columnwidth]{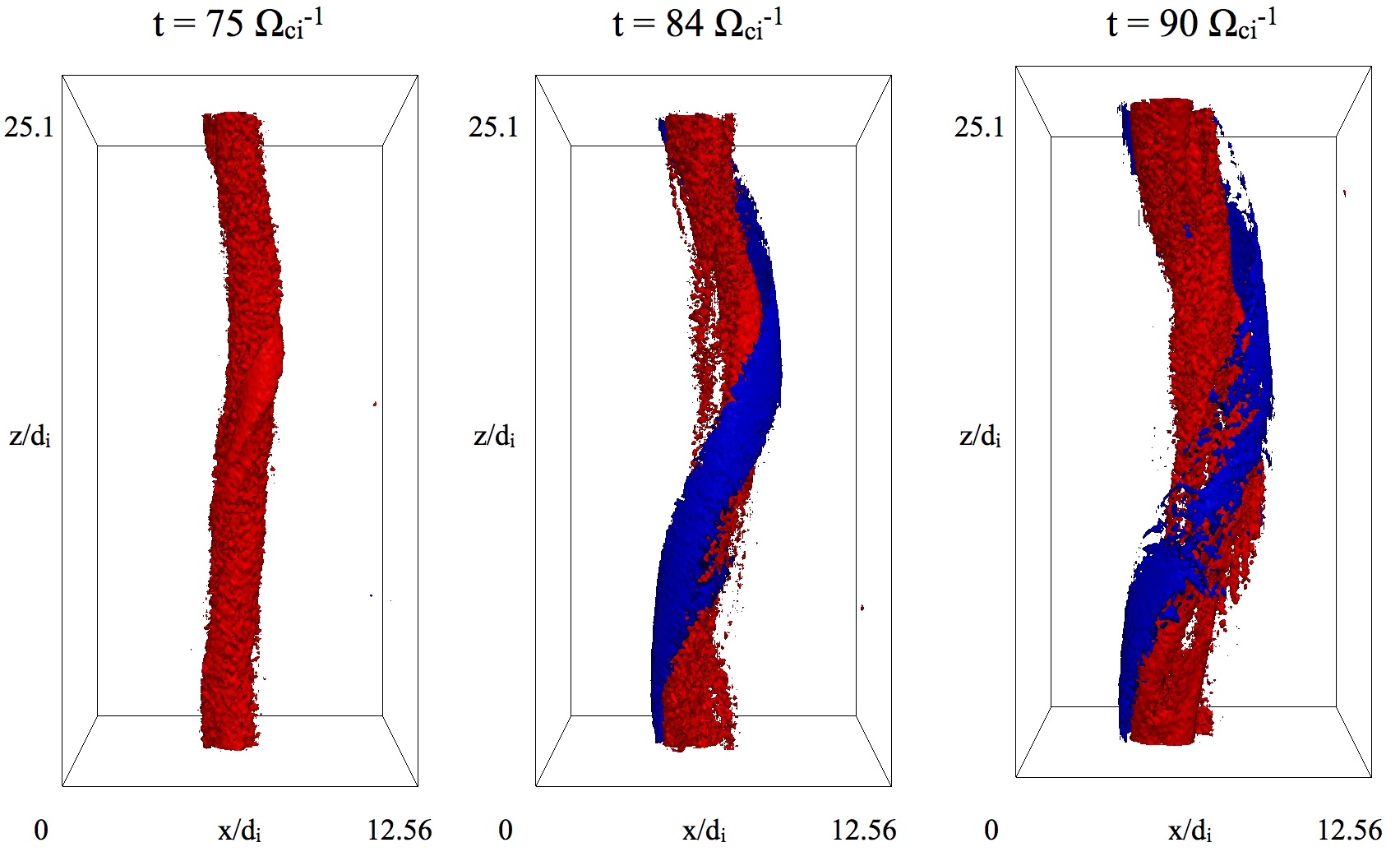}
\caption{Isocontour of the axial component of the electron current $J_{ez} =   0.005  \ e n_0 c$ (red color) and $J_{ez} = - 0.005  \ e n_0 c$ (blue color) at times $t= 75 \ \Omega_{ci}^{-1}, 84 \ \Omega_{ci}^{-1}, 90\ \Omega_{ci}^{-1}$.}
\label{Jz}
\end{figure}

Figure 9 presents an isosurface of the electric field component parallel to local magnetic field. The red and blue isosurfaces are for $E_{//} = \pm 0.04 \ B_0 V_A/c$ and an isosurface of the axial component of the electron current is superimposed to visualize the flux rope. Several bipolar parallel electric field structures develop along the reconnection line in proximity to the interface between the electron currents flowing in the opposite directions.

\begin{figure}[ht]
\includegraphics[width=1 \columnwidth]{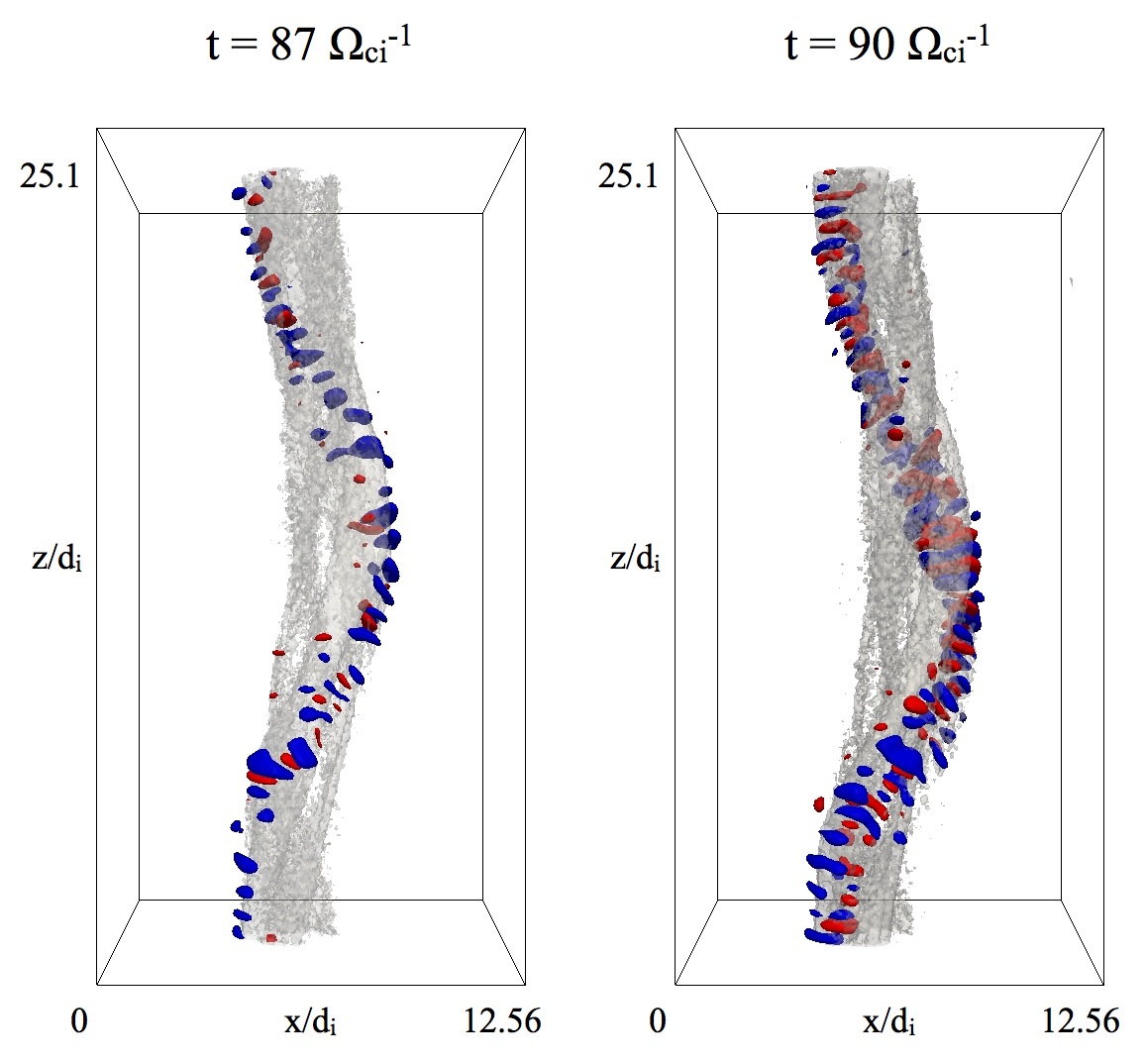}
\caption{Isosurface of the parallel electric field for $E_{//} = \pm 0.04 \  B_0 V_A/c$ with superimposed isosurface of the axial component of the electron current $J_{ez} = 0.005 \ e n_0 c$ at times $t = 87 \ \Omega_{ci}^{-1},  90 \ \Omega_{ci}^{-1}$. Several bipolar electric field structures develop in proximity of the reconnection region.}
\label{Epar}
\end{figure}


\section{Discussion and Conclusions}
In this study, the signatures of secondary collisionless magnetic reconnection triggered by the kink of an infinite flux rope have been identified. The kink instability of the flux rope has been simulated with a fully kinetic and fully electromagnetic PIC method, allowing us to retain the kinetic effects during magnetic reconnection. This is important when modeling magnetic reconnection, because fluid models can not provide a full description of collisionless phenomena.

As a first step of our study, the reconnection regions have been identified using the auxiliary magnetic field $\mathbf{B}^*$. Several signatures of secondary magnetic reconnection have been found on a plane perpendicular to the flux rope axis. Regions of enhanced and depleted density (with peak density 50 \% the background density) form a quadrupole structure in proximity of the reconnection region. The regions with depleted density are called "density cavities" in the literature and electron currents flow along them. The ion flow develops in large part on the $(x,y)$ plane. The decoupling of electron and ion dynamics on the $(x,y)$ plane generates an Hall magnetic field. On this plane, the Hall magnetic field has a bipolar structure in the plane. The peak value of the Hall magnetic field is 20 \% the axial magnetic field $B_0$. The most intense electric fields develop perpendicularly to local magnetic field and mainly on the $(x,y)$ plane. In proximity of the reconnection region, an electric field in the $z$ direction forms as effect of magnetic reconnection. Its value is $0.1 \ B_0 V_A / c$ (panel b of Figure 6). This is the so-called "reconnection" inductive electric field. During the secondary magnetic reconnection, an electron current develops in the opposite direction of the initial electron current of the flux rope. The intensity of this current is approximately of the same intensity of the initial electron current of the flux rope. The bulk electron velocity is approximately $1.5 \ V_A$. We believe that the reverse current would render the kink more stable. The presence of opposite currents has been observed in the Reconnection Scaling eXperiment (RSX) device  \cite{furno2003reconnection, Intrator:2009}. The interface separating the two opposite electron currents is characterized by the presence of bipolar electric field structures. It is likely that these structures are caused by electron streaming instabilities along the reconnection line \cite{markidis:2013, lapenta2011bipolar}. Because of the presence of several bipolar electric fields and electron-scale instabilities, the reconnection region is in a turbulent state. The turbulence of the reconnection region might affect the reconnection rate by introducing an anomalous resistivity \cite{biskamp1986magnetic}.

This paper is an example of study of magnetic reconnection in a non trivial initial configuration. We found that the results are generally in a good agreement with the results obtained starting from simple initial configuration, such as the Harris configuration with a uniform background guide field $B_0$ \cite{Rogers:2003}. In the flux rope configuration, the Hall magnetic field shows a bipolar structure differently from the famous quadrupolar structure developing in the Harris sheet configuration \cite{drake2007fundamentals}. However, the peak value of the Hall magnetic field $B_z - B_0$ is approximately 20 \% the core magnetic field $B_0$ as in simulations starting from Harris sheet configuration \cite{Lapenta:2010,MarkidisJGR:2011}. The reconnection electric field is  $0.1 \ B_0 V_A / c$ in both configuration also \cite{drake2007fundamentals}. As in simulations of Harris with a guide field, intense electron flows develop along the cavity channel. However, these electron flows are approximately $1.5 \ V_A$ in the flux rope configuration, while they are $6-7 \ V_A $ in the Harris configuration with a $B_0$ guide field \cite{markidis2012three}. Bipolar parallel electric fields develop in both configurations with typical electric field peak values $0.05-0.1 \ B_0 V_A / c$ and $0.3 \ B_0 V_A / c$ for flux rope and Harris configurations respectively \cite{markidis2012three,lapenta2011bipolar}. In the case of Harris sheet configuration, the bipolar parallel electric field structure originate along the magnetic reconnection separatrices, while they originate along the reconnection line in the flux rope configuration.

In summary, the results of a three-dimensional PIC simulation, modeling magnetic reconnection during the kink instability of a flux rope, have been presented. The main signatures of secondary magnetic reconnection can be found on a plane perpendicular to the flux rope axis. A quadrupolar structure of enhanced and depleted density, a bipolar structure of the Hall field and a reconnection electric field develop in proximity of the reconnection region. The reconnection line is characterized by the presence of several parallel bipolar electric field structures. These results can be used to identify magnetic reconnection during the internal kink instability.

This study focused on the analysis of signature of magnetic reconnection during kink instability of a single infinite flux rope. Future extension of this work will be the study of magnetic reconnection signatures  with multiple flux ropes. In this case, magnetic reconnection is triggered by combined kink and coalescence instabilities \cite{Markidis:2012,markidis:2013} making the analysis of the signatures more challenging. In addition, it would be important to study reconnection in flux ropes with an anchored end instead of using periodic boundary condition in the $z$ direction. This would allow to model more realistically laboratory experiments  \cite{Intrator:2009} and several space and astrophysical events.

\section*{Acknowledgments}
The present work is supported by NASA MMS Grant NNX08AO84G. Additional support for the KTH team is provided by the European CommissionÕs Seventh Framework Programme under the grant agreement no.~287703 (CRESTA, cresta-project.eu). Additional support for the KULeuven team is provided by the Onderzoekfonds KU Leuven (Research Fund KU Leuven) and by the European CommissionÕs Seventh Framework Programme (FP7/2007-2013) under the grant agreement no.~263340 (SWIFF project, www.swiff.eu). 
\section*{References}
\bibliographystyle{unsrt}

\end{document}